\documentclass{elsart}
\usepackage{epsfig}
\usepackage{amssymb}

\begin{document}
\begin{frontmatter}

\title{
Magnetic resonance as a channel of 
directed transmission of electromagnetic energy in animate nature.
}
\author{E.Ya. Fursa }
\address{Belarusian State University \\
Bobryiskaya 4, 220050, Minsk, Republic of Belarus }
\date{\today}
\maketitle

\begin{abstract}
The phenomenon of magnetic resonance (either NMR or ESR) is a responce of 
atomic (molecular) system to the external electromagnetic effect. Electrons 
and nuclei, which possess magnetic moment, are the "magnetic antennas" in the 
biosystem "human being". They are able to receive (radiate) electromagnetic 
energy selectively in coordinates $H$ (magnetic field), $\nu$ (frequency) and 
$\gamma$ (gyromagnetic ratio). Tuning of these antennas mirrors the state of
an atomic system and its environment. The sea of electromagnetic waves of both 
natural and artificial origin surrounds a "human being". It contains the set of 
frequencies, which fulfill the resonance condition in the geomagnetic field. 
This phenomenon is universal for a lot of nuclei and electrons, which 
participate in many vital biochemical reactions and processes. The above allows 
one to speak about the existence of a natural channel for energy (information) 
transfer using the radio-frequency range. This channel exists both inside a 
biosystem and for communication outside. The possible ways of realization of 
such a channel and trends of its application are discussed.
\end{abstract}
\end {frontmatter}

\newpage
\centerline{Contents}

\begin{table}[tbph]
\vspace{0.3cm}
\par
\begin{tabular}{ll}
{Introduction}& 3 \\ 
\hspace{1 cm}{\em Zeeman splitting of the levels. Magnetic resonance}& 3 \\ 
\hspace{1 cm}{\em Magnetic resonance from the other point of view}& 3 \\ 
\hspace{1 cm}{\em Principal concept}& 5 \\ 
{Value of resonance parameters for natural conditions}& 5 \\ 
\hspace{1 cm}{\em Resonance frequency} & 5 \\ 
\hspace{1 cm}{\em Energy of resonant interaction} & 7 \\ 
{Electromagnetic compatibility as a feature of environment}& 8 \\ 
{Connection of the observed phenomena with the magnetic resonance} & 13 \\ 
{Possible implementation mechanisms} & 14 \\ 
\hspace{1 cm}{\em About water and hydrogen bonds}& 14 \\ 
\hspace{1 cm}{\em Biophysics and cybernetics} & 16 \\ 
\hspace{1 cm}{\em Quantum-mechanical tunneling} & 17 \\ 
\hspace{1 cm}{\em Enzymatic catalysis} & 19 \\ 
{Symmetry in nature}& 20 \\ 
{Conclusion}& 23 
\end{tabular}
\end{table}

\newpage

\section{Introduction}
\subsection{Zeeman splitting of the levels. Magnetic resonance}

In the simplest case the eigenstates of either a nucleus or 
electron, corresponding to the different alignments of magnetic moment, 
are degenerated. Magnetic field causes splitting of these levels.
Similar electron levels, the set of magnetic (Zeeman) those allows
population inversion, amplification and generation. 
  
Let us consider an object placed to a magnetostatic field $H$ 
and irradiated by an electromagnetic field with the frequency $\nu$.
(Here $\nu$ corresponds to the condition 
\begin{equation}
h \nu = g \beta H,
\end{equation}
where $g$ and $\beta$ 
are the constants, describing the object.)
In these conditions the atomic system  (containing either electrons or nuclei, 
which possess magnetic moment) resonantly absorbs the energy of electromagnetic field.
The above phenomenon is called magnetic resonance, there are nuclear magnetic resonance (NMR) 
and electron spin resonance.

The magnetic resonance is used as a fine technique for study of 
matter structure, diagnostics and so on \cite{1}.

Modern NMR-imagers let one to observe resonant absorption of electromagnetic energy by every 
part of human body.

\subsection{Magnetic resonance from the other point of view}

Let us consider the magnetic resonance from the other point of view.
The objects we are using to observe either nuclear magnetic resonance (NMR) 
or electron spin resonance phenomenon possess natural permanent channel for 
electromagnetic energy absorption (radiation).
We observe it without considering it's destination, purpose and manifestation in different
phenomena and processes, first of all, in biological systems.
To realize such a channel the following conditions are necessary:

1. Biosystem (or it's part) contains nuclei or electrons, 
possessing magnetic moment due to unpaired or free electrons, radicals, 
paramagnetic or ferromagnetic substances

2. External magnetic field, splitting energy levels.

3. Electromagnetic radiation with the frequency corresponding to resonant condition (1).

Our environment possesses al the necessary conditions for operation of such energy channel:

1. A human body includes 70$\%$ of water. Concentration of nuclear spins (protons) in water is about
$5.1022 см^{-3}$. Nuclei 19F, 13C, 31P, 11В, 17О, 15N, 59Со also possess magnetic moment.
They are contained in a human body in different amounts. Besides, 
electron and proton transfer supports
all vital biochemical reactions, namely: redox reactions, free radical reactions, catalysis and so on.

2. Constant geomagnetic field of the Earth (0.7 gauss $\pm$ fluctuations caused by 
internal and external sources).

3. Natural electromagnetic background (lightnings, synchrotron radiation, electromagnetic waves 
radiated by cells of living organisms and so on) and artificial electromagnetic background 
(tele and radio communication, power engineering etc.)

Thus, conclusion issues: biosystem "HUMAN BEING" can not avoid immediate energy exchange with environment.
The frequency of such exchange is determined by the magnetic resonance frequency.

\subsection{Principal concept} 

A person can measure out the radiation effect upon organism
or it's parts and systems and utilize it in medical practice.
Practically, the full range of electromagnetic waves is used:
radio-frequency, visible, X-ray and $\gamma$ radiation.
Usually, it means the integral action upon either organism as a whole or it's part 
by heating.
Conventional methods and facilities do not allow selective controlling of biochemical 
processes in human body by the direct action on the definite molecules and nuclei, which are
responsible for pathologic processes.

In certain conditions, one can consider electrons and nuclei, possessing magnetic moment, as
"magnetic phased receiving arrays" in "HUMAN"-biosystem. Such arrays can receive (radiate) 
electromagnetic energy selectively in coordinates $H,~\nu,~\gamma$ 
(where $\gamma$ is the gyromagnetic ratio).
Tuning of such an array mirrors the state of the atomic system and its environment.

Ocean of electromagnetic waves surrounds and permeates a person. It includes frequencies from   
0.001 to 1000 MHz, which meet the condition of magnetic resonance in the magnetic field of the Earth for
several nuclei and electrons, participating in all vital biochemical reactions and processes.
The principal concept is:
both nuclear magnetic resonance(NMR) for protons (and other nuclei) 
and electron spin resonance (ESR) for free electrons provide
physically true correlation between a human being and environment. It is power, 
aura with high selectivity. 
It is true 
channel of energy interaction, which controls biochemical processes: speed and direction of response, 
which can be measured
and checked by physical means.
       
\section{Value of resonance parameters for natural conditions.}
\subsection{Resonance frequency.}

Frequencies corresponding the transitions between magnetic levels 
are in a wide range from audio frequencies to
long-wave microwaves:
\[
\nu=\frac{E_1-E_2}{h}=\frac{\gamma}{2 \pi}H
\]
where $\frac{\gamma}{2 \pi}$ is the proportionality factor.
For H nucleus $\frac{\gamma}{2 \pi}=4.26~kHz~per~gauss$ and for 
electron spin $\frac{\gamma}{2 \pi}=2.8~kHz~per~ gauss$

Nuclear magnetic resonance between Zeeman levels can be easily observed in the range 1-100 MHz.
Intense NMR signals (as well as spin echo signals)  can be obtained for hydrogen nuclei in ordinary water.
In the magnetic field of the Earth (0.5 gauss) the NMR frequency for protons is
\[
\nu_{{\small NMR}}=4.26~kHz/gauss \dot 0.5~gauss = 2.13~kHz
\]
The ESR frequency in the magnetic field of the Earth for free electrons is
\[
\nu_{{\small ESR}}=2.8~MHz/gauss \dot 0.5~gauss = 1.4~MHz
\]

In an actual case nuclei and electrons are contained in atoms and molecules, thus the interaction 
picture is more complicated \cite{2}.
The spin of a nucleus can undergo the action of internal local fields, 
which are produced by the spins of either 
neighbor atoms or unpaired electrons. In turn, these fields can 
be modulated by the rotation of molecules and so on.
If taking into consideration both the internal magnetic fields and phenomena of chemical insulation, 
one can obtain very
complicated spectrum for both NMR and ESR.
Nuclei possessing quadrupole moment are aligned by the electrical fields, produced by 
the valence electrons in a molecule.
This also causes variations in magnetic resonance spectrum and demonstrates 
the possibility of interaction with external
magnetic fields over this channel.
Though, very weak magneto-optical interactions are possible over either direct or inverse Faraday effect.

Life-time for the certain state of the system and the amount of the accepted energy 
are determined by its environment.
The observed phenomena of cross-relaxation, dynamical polarization of nuclei and others
demonstrate complexity of interaction of magnetic spin states.
Saturation of electron spin resonance of proper paramagnetic material at high frequency increases
the intensity of low frequency NMR signal several hundred times. And in some cases it can even change 
the sign of absorption i.e. radio-frequency radiation appears.

Water with paramagnetic additions can be used as an actuating fluid for paramagnetic amplification and 
generation of electromagnetic waves.
For example, as far as 1960 it was developed the maser with the actuating fluid, including water 
and paramagnetic salt $(SO_3)_2~NOK_2$
The maser is easily self-excited at room temperature in the magnetic field of the 
Earth and radiates with the frequency 2000 Hz at pumping frequency 55MHz \cite{3}.

In the absence of magnetic field the contact interaction (isotropic hyperfine interaction 
of electron and proton) mostly contributes to the energy of a hydrogen atom. The energy of
a nuclear moment in the magnetic field, produced by the electron spin, corresponds this 
type of interaction.
ESR spectra includes only one spectral line(1420 MHz).This frequency slightly differs from that
for free electron.
NMR of the hydrogen atom become complicated by the influence of electron spin.
It is caused by the fact that the time of spin relaxation of electron in some cases is small.
Hyperfine interaction yields anomalous shift of NMR frequency and also can cause NMR line spreading.
The transition probability for NMR is 10 times less, than the transition probability for ESR. 
The magnetic field of
nuclear spins can strongly affect valence electron.
As a result the barrier of chemical reaction decreases.
All the above indicates that this phenomenon could be used to control the speed of radical processes.

It should be particularly emphasized, that magnetic resonance as a natural phenomenon can not be
identified with the spectrometer output, because every spectrometer has sensitivity and
resolution limits.

A lot of resonance lines can not be observed due to 
either considerable spread of electrophysical properties of the samples or
high speed of resonance processes.
Actual spectra is much more rich in the absorption lines and dynamics, it also includes
emission lines.

It should be supposed the presence of local magnetic resonance phenomena on the boundary of 
biomolecules, cells
and membranes, which act as regulators or codes (key + lock) due to the coherent action of wave functions of
protons, electrons,...
Picture of resonance phenomena (spectra) for any biological object can be extremely complicated and dynamic.
Using the cybernetics language, this picture is data-capacious and can be used as a code system
for communication, control, action and so on.

\subsection{Energy of resonant interaction}
  
Orbital energy levels differ by tens of eV, while for Zeeman those the interlevel distance is about 5-10 eV.
The energy of magnetic resonance for $H=3000~gauss$ is: $E=4 \dot 10^{-3}$ kJ/gram-molecule
For comparison, the energy of covalent bonds is 170-630 kJ/gram-molecule, the energy of hydrogen bonds is
12.6-33.6 kJ/gram-molecule, the energy of Van der Waals interaction is 4-13 kJ/gram-molecule and the thermal
energy of a molecule (kT) at 300$\circ$K is 2.52 kJ/gram-molecule.
It means that the energy of resonance interaction is negligibly small for the 
energy scale of chemical reactions. The latter fact is, apparently, the main reason 
why magnetic resonance is not considered as an active agent for biochemical reactions.

Electrical forces between neighbor atoms are usually exceed magnetic those, because charges inside an atom
move with the speed, which is significantly less than the speed of light.
Presence of either ion or asymmetric charge distribution inside a molecule yield the electric field, 
which is mainly responsible for attractive forces in many materials. 
This field is more intensive
then the external field usually feasible in the laboratory. 
While the internal magnetic fields, caused by the motion
of charges inside the matter, are usually weaker in comparison with the external those. The action of 
internal and
external fields can be compared only in ferromagnetics. The higher the external field is, 
the more the magnetic levels are splitted and, therefore, the more energy the sample absorbs 
from the outside.
The question arises whether it is to good of living organism or to it's detriment.
Transitions between the levels occur at simultaneous action of a uniform magnetic field
and the magnetic component of an oscillating electromagnetic field on atoms. Interaction of an atom with
the oscillating electromagnetic field is realized through oscillation of mean magnetic moment of atoms
(nuclei) rather than electric dipole moment. 
Such magnetic interaction is less intensive than electric that,
but for microwave or less frequencies, the more intensive (several oders higher) oscillating fields 
can be applied, then for visible range frequencies. Moreover, frequencies in radio- and microwave
ranges can be tuned with high accuracy.

\section{Electromagnetic compatibility as a feature of environment}
Human being is the child of the Earth. 
Environment accompanies his existence from birthday to death. The energy of space is his food,
obtained indirectly from the organic food and directly as electromagnetic energy through heat 
and light.

Every physical object (including biological ones) continuously radiates and absorbs the energy of 
electromagnetic waves in wide frequency range in the form of rigorously dosated quanta with 
definite (resonance) frequencies for radiating and absorbing atomic (molecular) system.
Any violation in the transmitter-receiver system yields consequences for organism.
Let us consider two physical realities characterizing environment of human being, they are:
quasistationary geomagnetic field  and endogenous and man-caused electromagnetic background.
A lot of natural phenomena in biosphere with parameters of our interest are not explained yet, besides 
they are supposed to be caused by the presence of magnetic fields and electromagnetic radiation 
in environment.

{\bf The magnetic field of the Earth.} It's magnitude is small (0.7 gauss) and it does not 
 give  apparent concern for inhabitants. What is it's role in the planet life?
A beam of charged particles coming from the space is deflected and braked by the Earth field.
Synchrotron radiation, which appears here, reaches the Earth surface and brings some 
changes in the life of biosphere.
Magnetic field of the Earth undergoes perturbations about 0.01 gauss, also 
there are day-to-day variations of magnetizing force of the Earth due to lunar (10-4 gauss)
and solar (2.10-3 gauss) fluxes.
The response of all alive to these perturbations is also well known:
oscillations of birth and death rates, mental disorders, changes of crop capacity and so on
\cite{4}.
Geomagnetic field influences the symmetry of biological objects and their functional properties.
Change of geomagnetic field (GMF) changes rhythm of different variations in biological objects:
day-to-day variations, seasonal behavior, annual oscillations.
Universality of GMF action on living organisms with different levels of arrangement has been
noted by many authors
\cite{5}.
There are some proofs that orientation of a body with respect to magnetic poles of the Earth
can influence the behavior and physiology of people, animals and bacteria.
On the one hand, the GMF protect all the living from the baneful space radiation, but, on the other
hand experimental data shows that full shielding from the magnetic field yields degeneracy of 
some kinds of animals as early as in the fourth generation.
The long-term stay in hypo-magnetic conditions arouses delay in cell differentiation and 
appearance of mutant cells. Magnetic field changes the speed of chemical reactions in systems
with free radicals, for example, in liquid crystals (the basis of intracellular structure).
Therefore, it influences on the biologic permeability of membranes.

There is paradoxical situation: human senses do not respond to constant field of the Earth,
while it's absence yields degeneracy and death. "The life itself comes into an organism
and evolves over geomagnetic field."
Common sense suggests that weak magnetic field can not change something in bioprocesses essentially.
It provides something similar action of "inner grid of triode", when small change of voltage drive
causes big effect, multiply amplified through resonance (!) ... or sets the direction of a
process (rotation) as a whole i.e. symmetry.
Constant magnetic fields penetrate organism structures "dissembling" them. There are no losses 
in constant magnetic field. Energy dissipation appears only at presence of variable 
(changing with time)
magnetic fields.
From this is often erroneously conclude that permanent magnet does not provide ill effect, because
it produces stationary, static magnetic fields, which have not electric component.
Actually, considering the fact, that in constant magnetic field there are no losses, 
logical reasoning does not allow one to suppose direct action of the field on an object.
The magnetic field bring the system out of a "neutral" state, removes degeneration and splits levels.
Being coupled with an electromagnetic wave of certain frequency it can act on the object 
through resonance absorption.
This action is, probably, essential for living matter, may be even global.
"The geomagnetic field, just as gravitational that, is  all-penetrating and in-depth fact,
which, owing to its properties, inevitably ought to influence on processes on the Earth 
and its environment."
One can suppose that just the small addition to the magnetic field of the Earth, provided
by the Moon (Sun), sets (disrupts) the rhythm of biological clocks.
Sun bursts are the main source of interplanetary disturbances and intense beams of 
charged particles, principally, electrons with the energy 10-100 keV and protons with
the energy higher than hundreds of MeV. In 40 or 50 minutes after the Sun burst 
the horizontal component the magnetic field of the Earth $\Delta B$ increases
tens and hundreds of $\gamma$ ($1 \gamma=1 nanotesla$).
Ionospheric violations and breaking of radio communication occur.
After 0.5-5 hours $\Delta B$ decreases sharply by 40-400$\gamma$.
It continues from several hours to two or three days.
Dimensions and shape of magnetosphere of the Earth occur essentially changed.
How many biological triggers became turned on after such burst?
Unicellular organism possesses biological clock with endogenous rhythm.
It is supposed that self-sustained oscillation system of biochemical reactions inside the cell
can produces such a clock.
Change of any parameter of environment can indirectly manage the process due to feedback (
without direct action on processes in organism). For example, the sex of a chrysalis of the 
silkworm can be determined by proper temperature. In this case the change of a parameter is 
important rather than its absolute value.
The sensitivity of living organisms to the external excitation is very high. This fact
is experimentally ascertained. Electrophysical processes in human nervous system
occur at signal amplitude 10 V, eye responds single light quantum and limit of
acoustic perception is comparable with thermal noise.
But there no experiments considering influence of the Earth field on living organisms.
All the theories of magneto-biological effects, based on the processes of paramagnetic and
diamagnetic magnetization are useless in principle.
The orienting possibility of magnetic atoms of the living organism in the magnetic field
with the field strength 10 oersted (by Dorfman estimation) is two orders less then thermal action,
 i.e. negligible
 \cite{18}.
 Therefore, to summarize, the experiments studying influence of magnetic field action on living
 organisms and plants requires strong theoretical ideas and scientific suppositions.

{\bf Electromagnetic background}. The Sun and the Galaxy are the sources of 
electromagnetic radiation of wide spectra incident the Earth. It includes radiation of different types:
synchrotron and relic radiation (Fig.1).
Cosmic rays
(protons, nuclei of heavy atoms, electrons) 
are the source of synchrotron radiation.
Electrons of initial cosmic rays moving in interstellar magnetic fields radiate radio-waves.
Synchrotron radio-wave radiation of metagalaxy sources is linearly polarized and 
polarization level usually does not exceed several percents.

\begin{figure}[htbp]
\epsfysize=4.0 cm
\centerline{\epsfbox{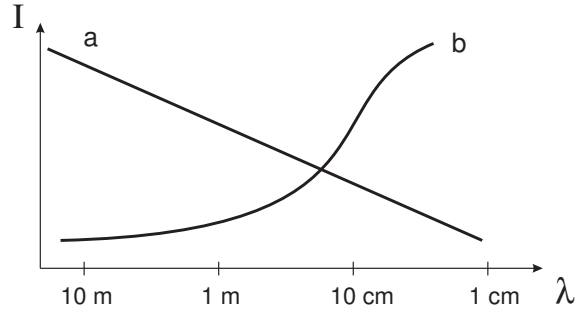}}
\caption{a). Synchrotron radiation of galaxy, b). relic radiation \cite{6}.}
\label{f1}
\end{figure}

In interstellar conditions the neutral hydrogen atoms radiate spectral line with the
wavelength 21 cm, hydroxyl, water vapor and oxide of carbon radiate
21 cm, 1.35 cm and 2.64 cm, respectively.
Process of intense radiation in spectral lines of hydroxyl and water vapor is
similar to operation of powerful natural maser.

Relic radiation was discovered in 1965. It is isotropic with spectrum, corresponding the radiation 
of the blackbody heated to 3 K. Density of radiation is 10 erg per cm. This is the highest 
energy density
in Galaxy.

A lot of Earth phenomena displays connection with solar activity.

Lightnings, being the sources of wide spectra of electromagnetic waves, played an important part
in transformation of life on the Earth.

The web of powerful transmission systems, operating in wide frequency range, winds around the globe.
This is one more source of man-caused radio-wave radiation background.                 
Every pocket radio-receiver can find out and detect the huge discrete set of radio-waves.

Human activity changes parameters of the Earth on the space scale.
I.S. Shklovsky evaluated that brightness temperature of the Earth as a planet was increased
during last 2 or 3 tens of years millions times more due to
operation of television, radars, broadcasting in metric wave-length range. Now it is about
several hundreds of million degrees, which exceeds brightness temperature of the Sun for this frequency
range in the periods of the absence of sun-spots
\cite{6}.

This fact should put us on guard. What is the reaction of living? 
What have changed around us due to this?

It is very difficult to discover influence of electromagnetic
field with small density on biological processes in the cells "in vivo",
particularly considering changes in ecological conditions.
Extensive and long-term observations are necessary.
Let us start from the simplest and consider electric processes in living organisms,
which are described by the perceptible and well-detected parameters: 
currents, potentials and so on.

Here there is a curious observation. Such prolate disease as arrhythmia (fibrillation)
was a rare event as recently as 70 years ago \cite{7}.
Scientists are speaking about epidemic. Sharp growth of the density of electromagnetic energy
in the environment causes the reaction of conductive system of the heart. It seems there are
no other explanations. But what is the action mechanism?
From this point of view one should draw attention to the neural and cerebrum diseases, because  
nerve cells are most sensitive to the action of electromagnetic fields in radio-frequency range.
A cell phone radiates high density of electromagnetic radiation in radio-frequency range (some systems
use frequency less than 900 MHz)
in the immediate vicinity of brain. What safety is expected? It is curious (at least I am curious)
in what generation of people the hereditary moronity or something similar become apparent?
Cell potential, conductivity of membranes and metabolism as a whole become strained.
Space flights make studies of the influence of the above factors on a human being more active.
Huge amount of experimental facts, which describe this action on biological objects depending 
on radiation frequency and intensity, is stored \cite{8}. However, there is no agreement in 
explanations of a diversity of the observed phenomena.
Attempts to explain injurious effect of electromagnetic fields on the vital functions of a 
human being yield to the necessity of consideration of  electromagnetic compatibility for
biosystems and environment \cite{9,10,11}

Actual environment is the net of cross-cut signals, where synergetic effects and new
interference low-frequency "constructions" can appear from two signals with higher frequencies
\cite{12}.

The influence of electromagnetic waves with mains frequencies (50, 60) Hz has been studied.
Prof. Volfgang Ludwig considers that fields oscillating with frequencies 3,  7.8 and  20 Hz
resound with organism structures and can activate them. We have already known that 
weak electromagnetic fields can strongly affect cell's growth and performance that, however, can not
be caused by some thermal exchange, because elsewise  thermodynamics laws are violated.   
Dr. Ross Edie shown in his investigations that physiological reaction can be provided only
at certain parameters of electromagnetic field, such as: amplitude, frequency modulation and so on.
Here there are recent results of Dr. Fabien Mamone: the impact by tone of the certain frequency during
15-20 minutes could activate healthy cells, while cancer cells are extended and breaked \cite{13}.

Presence of biorhythms is proven for many biosystems. They are explained by the
self-oscillating processes (hysteresis in cell structures), but no one has been
realized how these processes are coupled with external factors and what could synchronize
them.
There is no scientific explanation for unique abilities of extrasensory and telepathic individuals.
At the same time the existing scientific and experimental basis provides to simulate
human organism as a whole (or in part) as either generator (wave source) or receiver with tunable frequency 
and high selectivity. Tuning could be realized by volitional action, for example.
On the one hand it can explain distant information exchange. Such exchange can be easily 
realized between twins or close relatives, because their individual oscillations are close by their 
parameters. It often can be observed in life.
On the other hand this exchange supposes the existence of biological field or aura as a shield
of a biological object produced by evolution. Why a lot of human diseases become more intense at night?
What is the mysterious influence of moon-light? Here one should draw attention to the difference in
day and night radio-spectra. It is well-known that magnetic fields of the Earth
 on the day and night sides differ by the flux density (action of the solar wind).
 The frequency of the light incident on the Moon determines Moon transparency as a body, which possesses
finite electroconductivity.
At fast change of interplanetary magnetic field it appears a delay of reflected wave phase in 
"magnetic shadow" of the Moon.
It is possible that after reflection by the Moon
the part of polarized radio-radiation grows and it's resonance actions increases.
Or, may be,
the interference of forward and backward (reflected by the Moon) waves
excites holographic pictures and hallucinations
in the brain of  sleeping person.
Simultaneous consideration (comparison) of several factors, joined by the resonance condition can 
appear the key for understanding of many questions.
Analysis of spectra of magnetic resonance frequencies for the molecules and compound, which  
frequently occur in biological objects, and it's comparison with the background spectra including
lines of space radiation of Н, ОН, Н О and relic radiation can help to answer these questions.
The presence of polarized synchrotron radio-radiation and magnetic resonance in geomagnetic field 
requires particular attention.

Interaction is provided by coherent, phased and directed action of particle ensemble, 
first of all interaction of
protons by hydrogen bonds and interaction of electrons by 
free radical reactions. These fine effects are much more important in 
consideration
of interaction mechanism than the quantity of activating energy. 
One should look for contribution of magnetic resonance processes in self-oscillation 
processes in biosystems as well as in trigger and
tunneling effects.
I am
speaking about motion (transfer) of electrons and protons, but one should also
consider currents, which are produced by this motion, and, therefore,
other velocities and collective effect.
I am
speaking about activating energy, whereas one should consider resonance, phased phenomena,
for which small action causes effect (1/0), tunneling effect, amplification and generation.
For biological system the possibility of existence of a phenomenon similar to superconductivity at room temperature 
can not be 
excluded.

\section{Connection of the observed phenomena with the magnetic resonance}

Let us consider several biophysical phenomena from this point of view. There are some of them, which
can not be explained by modern theories and which can indicate (either directly or indirectly) 
connection with magnetic resonance for radio-frequencies.

And everywhere there are protons, paramagnetic atoms, free electrons and radicals i.e.
systems, which possess magnetic moment.

The magnetic field of the Earth is highly uniform and universal it determines the direction of
magnetic biasing for all biochemical processes in living matter on the Earth.
Magnetic moments of electrons and nuclei sensitively respond the slightest fluctuations 
of this field (similar compass needle).

Electromagnetic radiation, which participates in magnetic resonance interactions,
propagates and scatters in the Earth conditions well, it also provides minimal losses in biological
structures.

Almost all radiation sources near the Earth surface (both natural and artificial) operates 
in the above range.

Probability of spontaneous transitions is this range is small, therefore, only induced
absorption and radiation are significant; Einstein coefficients are about 4
orders less then those for electric dipole transitions.

The magnetic interaction is less intensive then electric that, but for low frequencies  
much more intensive oscillating magnetic fields can be used than for visible range.

The natural linewidth for magnetic resonance is very small and can be measured with high accuracy,
just as the phase. Electromagnetic field is considered in classical representation that allows to
speak about exact phasing (tuning) of large ensemble of particles.

A lot of processes in living matter (and abiocoen), being the processes in condensed
nonequilibrium systems, for example, enzymic catalysis, can not be explained  by the well-known
kinetic and static laws, based on the primitive heating and compression of material,
disordered thermal collisions i.e. statistic disorder. To all appearance, both
the cooperative action of particles with fine adjustment of phases of wave functions and
long range ordering are necessary for such phenomena.
This results in condition AND-AND rather than OR-OR, i.e. product rather than sum.
Cold nuclear fusion, which
caused a sensation ten years ago, is also pertinent to the above, in my opinion.
Nature is rich in unique solutions, passing obstacles around by elegant methods with
the smallest energy consumption and entropy decreasing. There are some samples:
covalent bonds in animate nature, quantum-mechanical tunneling and so on.

In many cases the activating energy is not the main factor in chemical transformations
and the explicable energy balance is absent.
Quantum-mechanical tunneling is the principal theoretical explanation.
The presence and influence of low-frequency modes of oscillations, which cause
smearing and swinging of electron levels can contribute the above.

Findings of influence of electromagnetic radiation on a human being (it is discussed above)
force to conclude the necessity of consideration of electromagnetic compatibility of 
the human being and its environment.
The lack of electromagnetic energy in an organism is a sign of homeostasis violation.

Physical processes in microcosm are the objects for directed physiological and biochemical
control, in other words they depend on conditions of their passing in organism and environment.
Biological mechanisms of development become apparent even for the lowest levels of life organization.

Magneto-optical effects and so on.

\section{Possible implementation mechanisms}
\subsection{About water and hydrogen bonds}
 
 Different types of forces participate in molecules interaction with each other.
 The particular type of intermolecular interaction, being of our interest, is the
hydrogen bond (H-bond). It is weak, the energy, which releases at H-bond forming is about
0.1 eV.
For example, the change of strong covalent bond requires the energy higher by a factor of 20.
It means that sufficient energy should be directed to the molecule for its activation.
The living organism can not solve this problem by the increase of temperature, because the
disordered thermal motion could change something unrequired.
The chemical reactions, either producing or destroying the covalent bonds in a cell,
occur due to participation of the specific catalyst, which are called enzymes.

In particular, the H-bond plays the key role in biochemistry due to its weakness.
Suffice it to say that the bond between two links in a DNA molecule and a secondary structure (conformation)
of proteins are formed by the H-bond. Since NMR directly acts on a proton, so the chemical processes using
H-bonds should be considered.
  
Speaking about the global synchronizing action of the geomagnetic field on biological and
physical-chemical reactions, one should put a question about the reason of 
such versatile action. The change of water properties under the action of
the geomagnetic field can be the reason. Because water, which is the common component in the 
reactions in living and abiocoen systems, is the only substance, which can determine 
the versatility of the geomagnetic field
action on the living organisms.

At the room temperature water should become gas, but the presence of hydrogen bonds displaces
the boiling-point of water to the anomalously high value 373.15 К at the atmosphere pressure.
The activation energy (for H-bond break) is about 20 kJ per gram-molecule at the temperature
298K.
Molecules of water undergo oscillatory motion near the equilibrium position with the typical
oscillation
period about
$2.7 \cdot 10^{-12}~s$.
The major properties of water are not quite clear \cite{14}.

Water directly participates in forming of structure of the most important biomolecules,
as well as in processes of self-assembly of complicated permolecular structures.
Small changes in quantity and state of relatively small fraction of water molecules, which form
hydrated layer of macromolecule, yield sharp changes in
thermodynamic and relaxation parameters of the liquor as a whole.
Fast process involves all fractions of water, including molecules
lying in the internal layers of protein.
The above makes the system "biopolymer-water" a uniform cooperative system,
in which all changes in the state of either solvent or macromolecule are
interconnected and correlated.
The huge and irreplaceable role of water in animate nature suggests that
the main destination of water is more important. It is
informational basis of biological life in the Universe.

H-bonds stabilize the secondary structure of polypeptide chains. Bond is realized
between hydrogen atom, which is connected to one molecule by chemical bond,
and electronegative atoms O, N, F, Cl, which are usually pertinent to the other
molecule.
The nature of hydrogen bond is very complicated and can not be explained by
the electrostatic attraction only, while it contributes the most in the
bonding energy.
Studies showed that potential function of H-bond can be represented by a curve with two minima 
localized near electronegative atoms. Proton tunneling is possible between them.

\subsection{Biophysics and cybernetics}

Biosystems can operate in several stable stationary states.
The main peculiarity of biosystems is their ability to switch 
between operation regimes. The ability of biosystems to switching (trigger effect) becomes
a precondition to the tissue differentiation (genetic trigger). Trigger switching is possible by
two ways:

1. the specific manner - by the sharp change of the variable quantity itself

or

2. the nonspecific (parametric) manner - the parameters of a system (for example, change of T, pH and so on) 
are affected instead of variables.

Transition from one stable state to another can be realized by different ways
according to the direction of parameter change. Here we can see nothing else than hysteresis, which, in
its part, can explain the presence of dynamic store inherent in all biological objects.

The geomagnetic field influences on the rhythm of all the processes in the living organism and biosphere as a whole.
Daily rhythm of radiation sensitivity (i.e. change of efficiency of radiating by ionizing radiation within a day)
is marked by radio-biologists. It is not clear yet, what is the timing element (the trigger) \cite{15}

The trigger properties of enzymatic systems play the key role in control of intracellular processes of metabolism.
The trigger properties of transport systems (for example, transport of liquor trough the
porous membrane) are well known.
The conformational state of an enzyme changes (switches over) at some critical concentration of
a substrate (or reaction product). This yields the sharp reduction of enzyme activity and, therefore,
slows down of the chemical process.

Nonlinear kinetics of the most important biochemical processes causes the presence of 
continuous periodic oscillations (self-oscillations) in biosystems along with trigger regimes and
hysteresis. The amplitude of the above oscillations is determined by the properties of the system itself
rather than the initial conditions.
An illustrative example of such reaction is the reaction of Belousov-Zhabotinsky.

Real biosystems are exposed to the infinite number of the accidental external and internal impacts,
but in the stable operation range the dynamic character of the system is retained.

There is a cybernetics approach to the life problem (Lyapunov's theory). Lyapunov describes the life as
"the highly stable state of a matter", which uses the information coded by the state of single molecules
for the development of "the  retaining reactions". To change the state of such a molecule the energy
of the disordered thermal motion of a biosystem is insufficient. The theory supposes the presence of a 
channel for communication with the outer world.
Information about external action is perceived by matter as coded signals.
A physical process can be the substantial realization of this signal.
But nobody have disclosed this process except for hard radiation, which is suspect of all the faults of
mutations and natural selection.

Finally, it should be mentioned that one of four conditions necessary for appearance of the processes of
spatially-temporal self-organization is the cooperativity of microscopical processes in the system. 
The thermodynamically opened medium, nonlinear system and minimal amplitude of interaction
are also required.
Modern thermodynamic theories describing active molecule-carriers ("molecular cars")
state that the speed of transport processes and energy transformation in the
system depends on the particular molecular-kinetic mechanisms of cooperative interactions.

\subsection{Quantum-mechanical tunneling}
In a number of cases the biochemical reactions go even when the necessary content of activation energy 
is absent. Electron can be transported at relatively large distances independently of translational motion
of molecules of electron donor and acceptor. This distinguishes the above processes from oxidation-reduction 
reactions in liquor. It is supposed that quantum-mechanical tunneling (QMT) takes place here.

The idea of tunneling transport of electron between single protein molecule-carriers separated by
the energy barrier can be applied to explain a lot of important phenomena in chemistry and biology.

As far back as in 1965 Levdin \cite{16} suggested that genetic defects, ageing and tumor growth can
appear due to QMT.
Genetic information is coded in molecules, which have hydrogen bonds between 
bases
in the specific pairs of nucleotides.
The normal pairs of bases can undergo tautomeric transformations by tunneling of proton from one base to another
in the replication process, which results in the appearance of faults in genetic code.
The detailed calculations show that the potential barrier between two minima in the system founded by the pair 
of bases is not too high and wide to make tunneling  the rare event.
To argue for electron tunneling is the fact, that tunneling can occur even at low temperature (77K and lower).
In these conditions the translational motion of large protein molecules is slowed down and, therefore, conventional
physical and mechanical mechanism of reaction (colliding particles with the excess of kinetic energy) can not be 
realized with high speed. In a number of cases the activationless tunneling prevails at $T < 100K$
and the temperature dependences of such reactions displays the pronounced biphase character.
Electron transition includes either transport of electron or migration of the energy of
electron excitation between two states in the system of two discrete levels; it can be realized
only in the presence of dissipative processes in acceptor.
The implication is as follows:  the energy of electron excitation (its part) should be scattered by
the acceptor during the time of electron being there. This yields to the "detuning"
of levels resonance in acceptor and donor during the defined time (very short time) and back
transition from acceptor to donor becomes impossible.
The states D (donor) and A (acceptor) are separated by the potential barrier and in the absence of 
dissipation the quantum-mechanical oscillations between the above states happen.
If the final state (A) is quasi-stationary (the level is metastable) due to certain
dissipative processes and is described by the complex energy, then the process of transition 
from the initial to the final state is one-way i.e. tunneling takes place.
Dissipation of a part of the energy can occur either due to further tunneling of electron through the
barrier walls to the acceptor neighborhood or under the action of the environment, which interacts with electrons.
These yield the "bouncing" of the energy level and loss of coherence of wavefunction states due to the shift of
the oscillation phase.
And the most interesting is that spreading of electron levels in complex molecules can be reached
due to the motion of nuclei and vibronic interactions, which result in the loss of electron energy through
the oscillatory degrees of freedom.

Then the summary is as follows:
consideration of electron tunneling in real molecules requires to take into account the state of
nuclei, which motion changes the energy levels of electrons.
Quantum character of nucleus motion causes the uneven change of oscillation energy of a molecule.
Oscillatory quanta correspond the oscillatory sublevels. There are some indications that electron transport in proteins
in real processes is described both by the quantum accepting modes and the set of low-frequency modes
inherent in protein molecule. Coupling of the low-frequency modes can influence the forming of the accepted mode and the
complex A-D as a whole. Just the case, which, in particular, demonstrates the active role of 
protein environment, explains the great difference in temperature dependence
for electron transport to different objects in photosynthesis reaction centers \cite{17}.

Quantum-chemical investigations let us to display some new peculiarities in the motion of nuclei of particles
containing in a molecule.  Thus, a lot of minima were discovered on the potential surfaces, which are separated
by the relatively low potential barriers. Moreover, very high sensitivity of electron structure of molecules in
excited state to the change of nucleus configuration and small external excitations was discovered two.
To investigate the intermolecular interactions, quantum chemistry should consider the potential of interaction and
develop the models, which can let us to take the influence of environment into account by studying the properties 
of a molecule and mechanisms of elementary processes.

\subsection{Enzymatic catalysis}

Chemical catalysts and enzymes directly concerns the considered question.
The transport of electrons and protons is a basis of activity of many of catalyst systems.
In the first case the typical catalysts are transition metals and their compounds.
A free radical is formed at single-electron transition. In the second case the
substances, which can accept and deliver a proton are used as catalysts.
Amasing reactivity of the enzymes has no convincing theoretical explanation. There is
lack of some ties, coordination, speed and energy balance and so on.
   
In biochemistry there are known such models as
Fisher's "key-lock", Kochland's "hand-glove", "protein-machine" and so on.
The thermodynamic essence of the above models is as follows:
the free potential energy of binding of a substratum (sorbtion) by an enzyme should be used for
lowering of a barrier of free activation energy in the following chemical reaction.
But it is not clear yet, in what forms and by what mechanism the sorption energy can be stored in
protein globule and concentrated to attack the bond. To explain catalysis mechanism it is supposed that
enzyme's structure provides coherent propagation of fluctuating changes of conformation from the protein 
surface to its active center. This provides free energy change between macromolecule and its environment.
The multi-step transformation of a substratum in enzymatic catalysis is provided by the synchronous and cooperative
transformations in a uniform polyfunctional system.
Assumption of the appearance of a coherent state with a single mode of oscillation and dissipativeless
energy transport over the certain degrees of freedom of a protein globule to the active center is a
base for all dynamic models. But no one of the mentioned models can be considered as standard.
The question about energy transport concentration and balance remains pressing.
Proton (proton pump)is mentioned in increasing frequency. Proton is the strongest catalyst.
Moderate excess of protons in blood plasma (pH parameter) signifies a lot in pathophysiology and
clinical practice.
The ability of protonated hydrogen to come close to electronegative atoms is its advantage.
 
Hydrogen atom lacking in electron environment does not meet electromagnetic resistance during approach to
other atoms, which have an electronegative cloud. It transports the excitation energy or angular momentum
to molecule. It remains only to phase-in the system of enzyme-substratum (to set the key into the lock) at minimal action
energy. In NMR conditions all the wave-functions of protons can be phased-in, but this implies the other features and 
other speed.

\section{Symmetry in nature}

Let us consider one more interesting fact. Some media possess an amazing property:
while a light beam with plane polarization passes through it, the plane of light polarization rotates.
Such media are called optically active (or gyrotropic). There are media providing right (clockwise) and left
(counterclockwise) rotation (for an observer located near the light source). Optical activity is
explicated by both the medium and molecule structure. From this appears "right" and "left" molecules.
Spatial structure of such molecules has no mirror symmetry. They are enantiomers. The absence of
mirror symmetry is also called chirality.
In abiocoen "left" and "right" molecules are found with equal frequency, whereas 
the molecules without mirror symmetry are usually present in living organisms in only one enantiomer type.
Pasteur supposed that this is the boundary between abiocoen and animate nature.

Alpha-spiral, which determines the structure of protein molecules, is usually the right screw,
while the aminoacid molecules forming a protein are usually left ones.
The DNA duplex is also right molecule.
All the DNA turns are right, similar ordinary corkscrew.
In nature there are no mirror reflections with gene spirals counter swirling. 
Due to the absolute symmetry and absence of mirror reflections the whole information contained in genes
can not be confused.  
Viruses are the protein compounds. They also rotate to the right.
Exclusions are antibiotics - they rotate to the left. Maybe this causes their action.
Different molecules devoided of mirror symmetry are usually presented only by either right or left stereoisomers.
All the above testifies that a living organism is described by full-blown left-right asymmetry.
Plants, like other organisms, are separated to "left" and "right" and symmetrical (  D-, L-,S- )
enantiomorphs.
Nature (plants and animals) produces and assimilates the aminoacides of only one isomer and can not assimilate the
synthesized imitations.
A human being feeds on the stereomers, which correspond his asymmetry.
Pasteur mentioned that "asymmetric synthesis can occur in the presence of some natural asymmetric factor".
In laboratory conditions depending on the direction of rotation of
light polarization vector mainly either "left" or "right" form of some synthesizing organic compound appears
under the action of light with the circular polarization.
As Sun radiation has no appreciable circular polarized component, then such explanation for the asymmetry of
"living" molecules is considered improper.
16 types of aminoacides, which was discovered in the Australian meteorite (1969) equally had 
"left" and "right" asymmetry. 
equally. It is evident that space aminoacides have nonbiological (or rather unearthly) origin.
This signify that the solution of the above phenomena should be find on the Earth.
The reasons of the asymmetry in nature are not clear yet. It is still a question, why some 
of symmetries are violated,
while others are not. It is not even clear what of two questions is more important:
why some of symmetries are violated or why certain of them are ideal?
There are assumptions that the latter is only a chance during life incipience.
However, one should always remember that any symmetry is a property of certain laws of motion, 
rather than abstract space.

Discovery of parity violation in elementary particle physics exhibited that nature distinguishes 
"left" and "right" and the choice between them is not accidental. B.Ya. Zeldovich et al. paid their 
attention that parity violation should cause the difference in internal energies for a pair of optical 
isomers. In 1983 
S. Mazon and J. Trenter
demonstrated that weak interactions, which violate parity, 
stabilize L-isomer in comparison with D-isomer. It is important that stabilization energy is 
negligibly small and depends on the angles of internal rotation.
But in normal conditions it can provide about 10 additional L-macromolecules per 1
gram-molecule of racemate. It has not been clear yet, whether such weak effect could lead to
wide-ranging results or not.

Nowadays, it is stated that the trend of physical-chemical reactions in protein molecules depends on 
its orientation and this effect, by some way, is determined by the object asymmetry 
(either left or right), the phase of Sun cycle in the moment of study and the geomagnetic field.
The influence of the geomagnetic field is clearly recognized and not impeachable.
It is supposed that variability in bi-symmetrical forms of plants is also caused by the
geomagnetic field and through it by the Sun activity in the Earth conditions.
In the place, where pigmies live, there are many nanoid animals and plants.
The Japanese, who grow up in USA (especially in the west part), differ drastically by their
height, ratio of both face and body from the parents in Asia. Two twins were in the navy in different
places: the first was in Baltic region and the second was in the Far East. As a result they
have got the height difference about 4 cm. And famous acceleration ... Maybe there is a prompt
in the fact that prolate allergy for exotic fruits, is caused by the conditions of their growth
(different geographic latitude and climatic zone) characterized by other ratio in 
asymmetry molecules.

Symmetry problems play determinative role in modern physics and can be a guiding thread to unknowable.

The extremely mysterious fact is the gyrotropy of the most important tissues of living organisms.
Namely, they are formed by the chiral molecules, which predominantly are in one of two mirror forms.
The numbers of the right and left molecules in nature are usually equal on average (racemate).
On my view the magnetic resonance concerns with the question of asymmetry in nature.
Physicists know the effect of magnetic circular dichroism (the Faraday effect). The right and left
components of light with plane polarization interact with matter in different ways in the presence of
magnetic field, which direction coincides with the direction of light propagation.
All the substances are active in terms of magnetic circular dichroism. This testifies that magnetic
fields significantly change the structure of electron transitions and cause asymmetry. The energy of a
particle does not depend on its orientation in space.
Invariance with respect to spatial rotations is determined by the conserved quantity the angular momentum.

Appearance of the preferential direction (along the magnetic field) violates the initial symmetry
with respect to arbitrary rotations around any direction in space. Thus, the angular momentum $J$
can no longer be conserved.
Before the field appearance the energy levels of a system are $(2j+1)$-fold degenerated.
The field turning removes this degeneration (Zeeman splitting).
The constant field of the Earth determines such direction in space, so nobody can expect to find
pure symmetry in "sublunary world".
Therefore, all the biochemical processes and life in the Earth pass in asymmetry conditions caused by 
the magnetic field from their beginning to the end. And only fluctuations (fluctuations over value and directions) 
of the geomagnetic field periodically violate the normal practice.
 
Let us consider the classical model of the magnetic resonance. The imposed magnetic field $\overrightarrow{H}$
induces the angular momentum $\overrightarrow{\tau}$, which acts on the magnetic moment of either a nucleus or an electron 
$\overrightarrow{\mu}$.
This angular momentum can be expressed by the crossproduct
\[
\overrightarrow{\tau}=\overrightarrow{\mu} \times \overrightarrow{H}.
\]

\begin{figure}[htbp]
\epsfysize=4.0 cm
\centerline{\epsfbox{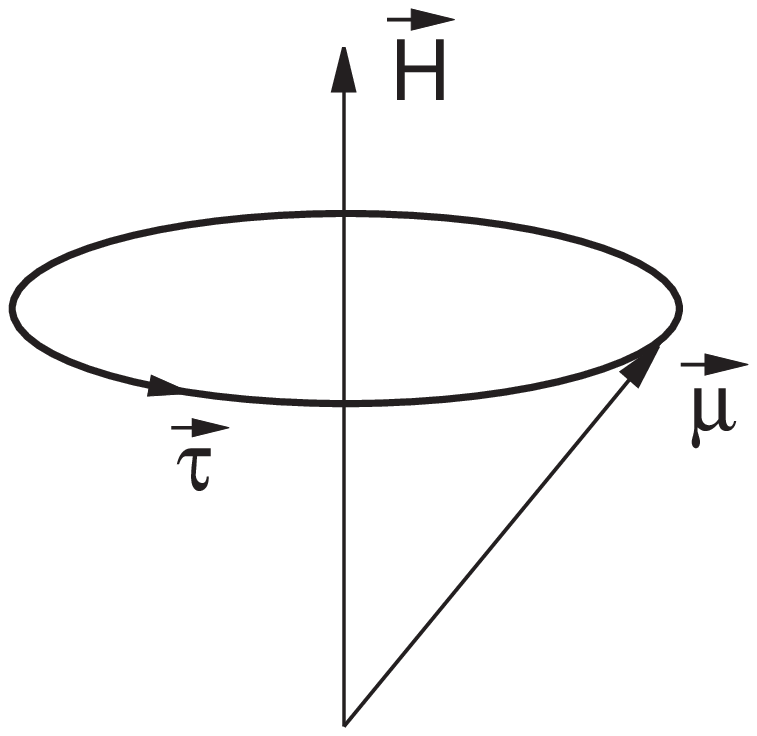}}
\caption{}
\label{f1}
\end{figure}

The magnetic field $\overrightarrow{H}$ makes the magnetic moment $\overrightarrow{\mu}$ 
to rotate (precess) around the direction
$\overrightarrow{H}$ with the frequency $\omega$
\[
\omega=\left| \gamma \right| H.
\]
Direction of rotation is determined by the sign of $\gamma$; $\gamma=\mu/\tau=g\beta/h$ is
the ratio, which is fixed for certain nucleus or atomic system.
In contrast to the most of nuclei, $\gamma$ for electrons is negative and their spins precess 
counterclockwise around the field $\overrightarrow{H}$.

It should be especially emphasized that transitions for NMR and ESR are induced by the magnetic component of 
circularly polarized radiofrequency radiation and the direction of precession coincides with the direction of
polarization rotation.

As one can see, the resonance binds rigidly both the frequency and direction of rotation. Maybe the tie of 
these phenomena hides here? There are common "characters in the play".
It could be supposed that just the magnetic resonance determines the direction of biochemical reactions depending 
on
orientation with respect to geomagnetic field, its value and stereoisomerism (either left or right).

\section{Conclusion}
During a century scientists have been considered chemical model of organization and operation of
molecules, tissues and organs of a body. This biochemical model should be expanded. Both 
electromagnetic and quantum processes should be included in it as they play a very important role 
in nature
self-organization.
  
Investigation of magnetic resonance participation in these processes could help to answer 
many of the above questions.
Even if magnetic resonance is not a reason of a lot of phenomena, it is an intermediary.
Just this side of the magnetic resonance effect I'd like to underline.        

Study and application of the magnetic resonance as a channel for electromagnetic energy transport allow 
to do the following:

1. To compare the spectra of magnetic resonance of the Earth for the most important (from the biological point of view)
molecules and compounds with the spectrum of electromagnetic background of biosphere. To calculate the spectral picture
of interaction of a human organism with the environment in the radio frequency range. First of all, to check
whether there is anomalous effect of frequencies 3Hz;  7.8Hz;  20 Hz,  3kHz,  2MHz,  1420MHz,  1700MHz,  22000 MHz;

2. To control the biochemical reactions in a human organism at atomic-molecular level selectively in coordinates
$H$, $\nu$, $g$. To rehabilitate (to correct) homeostasis violation in the exchange of electromagnetic energy with
the environment by the pumping  of the magnetic resonance energy to the required place, organ or system for
medical treatment; to activate healthy cells and inhibit ill those. Localization of the resonance conditions can 
be provided by several methods. The internal localization can be done by g-factor, by application of spin 
sones or contrasting substances. The external one can be provided by hardware (for example, the NMR-T code, which 
allows to distinguish T1 and T2 for healthy and ill cells).

3. To simulate and test biological possibility of a human organism (organs) as a generator of 
electromagnetic radiation (maser); to find possible antennas (points, chakras) in the body to explain
mechanism of telepathic connections, telekinesis, aura and so on.

4. Considering NMR ability to phase the wavefunctions of particle ensemble, to find out its role
in enzymatic catalysis, tunneling, molecule's asymmetry and mutations of DNA with H-bonds during the
replications and reparations in the operation of "proton pump" (as well as other reactions and compounds
with H-bonds);

5. To explain the influence of fluctuations of the geomagnetic field and Sun activity on the phenomena in
biosphere (including biorhythms)
in terms of magnetic resonance (MR);

6. To indicate the day and night difference in MR conditions for the Earth inhabitants due to the changes
in the geomagnetic field value, radio-background and its polarization as a possible reason of worsening of
the state of health of the most sensitive people. Magnetic resonance possibly could
be a transfer chain to realize the influence of the radiowaves reflected from the Moon on a human being
by creating holographic pictures in the brain of a sleeping person;

7. To prove the connection between the epidemic of heart arrhythmia in recent 50 years with the significant 
growth of radio brightness of the Earth in this period;

8. To find out the mechanism of influence of static electricity, which is accumulated on a body due to
synthetic clothes, cell phones and so on, on human organism;

9. To apply spine sones for "repairing" or destruction of cells (peculiar pole-hole digger) in ESR regime;

10. To examine the experiments and to develop a procedure for treatment of oncological deceases with 
electromagnetic radiation by the reversed NMR-T (NMR  imagers can be used for treatment: to localize the
pathology in conventional regime and then to radiate localized parts (nuclei));

11. To activate an organism by either application of the water "magnetized by NMR" or hemosorption
(NMR for water-containing systems and ESR for paramagnetic those).

\end{document}